\def\year{2020}\relax
\documentclass[letterpaper]{article} 
\usepackage{aaai20}  
\usepackage{times}  
\usepackage{helvet} 
\usepackage{courier}  
\usepackage[hyphens]{url}  
\usepackage{graphicx} 
\usepackage{graphicx}  
\usepackage{amssymb}
\usepackage{multirow}
\usepackage{amsmath}

\title{PHASEN: A Phase-and-Harmonics-Aware Speech Enhancement Network}
\author{
\Large \textbf{Dacheng Yin\textsuperscript{\rm 1}, Chong Luo\textsuperscript{\rm 2}, Zhiwei Xiong\textsuperscript{\rm 1}, and Wenjun Zeng\textsuperscript{\rm 2} }\\ \textsuperscript{\rm 1}University of Science and Technology of China\\ \textsuperscript{\rm 2}Microsoft Research Asia\\ 
ydc@mail.ustc.edu.cn,\ cluo@microsoft.com,\ zwxiong@ustc.edu.cn,\ wezeng@microsoft.com}

\begin{document}

\maketitle

\begin{abstract}
Time-frequency (T-F) domain masking is a mainstream approach for single-channel speech enhancement. Recently, focuses have been put to phase prediction in addition to amplitude prediction. In this paper, we propose a phase-and-harmonics-aware deep neural network (DNN), named PHASEN, for this task. Unlike previous methods which directly use a complex ideal ratio mask to supervise the DNN learning, we design a two-stream network, where amplitude stream and phase stream are dedicated to amplitude and phase prediction. We discover that the two streams should communicate with each other, and this is crucial to phase prediction. 
In addition, we propose frequency transformation blocks to catch long-range correlations along the frequency axis. Visualization shows that the learned transformation matrix spontaneously captures the harmonic correlation, which has been proven to be helpful for T-F spectrogram reconstruction. 
With these two innovations, PHASEN acquires the ability to handle detailed phase patterns and to utilize harmonic patterns, getting 1.76dB SDR improvement on AVSpeech + AudioSet dataset. It also achieves significant gains over Google's network on this dataset. On Voice Bank + DEMAND dataset, PHASEN outperforms previous methods by a large margin on four metrics.
\end{abstract}

\section{Introduction}
Single-channel speech noise reduction aims at separating the clean speech from a noise-corrupted speech signal. Existing methods can be classified into two categories according to the signal domain they work on. The time domain methods directly operate on the one-dimensional (1D) raw waveform of speech signals, while the time-frequency (T-F) domain methods manipulate the two-dimensional (2D) speech spectrogram. Mainstream methods in the second category formulate the speech noise reduction problem as to predict a T-F mask over the input spectrogram. Early T-F masking methods only try to recover the amplitude of the target speech. When the importance of phase information was recognized, complex ideal ratio mask (cIRM) \cite{williamson2016complex} was proposed aiming at faithfully recovering the complex T-F spectrogram. 

\begin{figure}[t]
\centering
\includegraphics[width=\columnwidth]{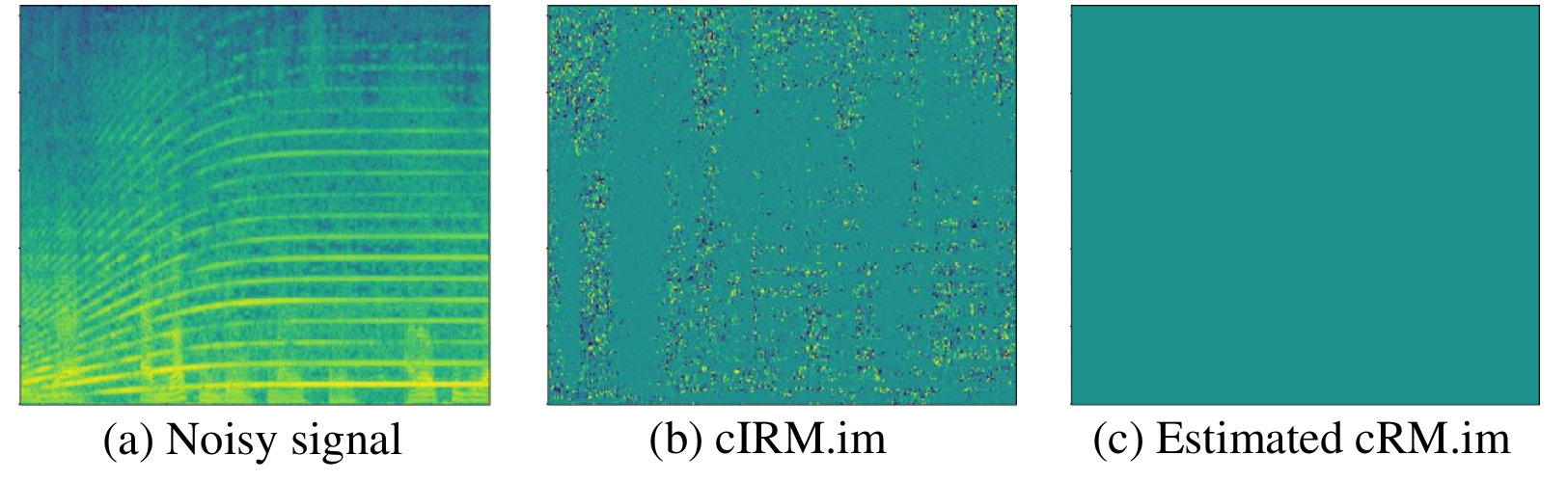}
\caption{Straightforward cIRM estimation does not achieve desired results. Although the imaginary part of the cIRM, as shown in (b), contains much information, that of a predicted cRM, as shown in (c), is almost zero.}
\label{fig:cIRM_artifacts}
\end{figure}

Williamson et al. \cite{williamson2016complex} observed that, in Cartesian coordinates, structure exists in both real and imaginary components of the cIRM, so they designed deep neural network (DNN)-based methods to estimate the real and imaginary parts of cIRM. However, in our evaluations of a modern DNN-based cIRM estimation method \cite{ephrat2018looking}, we find that simply changing the training target to cIRM did not generate desired prediction results. Fig.\ref{fig:cIRM_artifacts}(a) shows the amplitude of the noisy signal where the stripe pattern is caused by noise. Fig.\ref{fig:cIRM_artifacts}(b) and (c) show the imaginary parts of the ideal mask and the estimated mask, respectively. Surprisingly, Fig.\ref{fig:cIRM_artifacts}(c) is almost zero, meaning that the estimated cIRM is downgraded to IRM. In another word, the phase information is not recovered at all. 

This observation motivates us to design a novel architecture to improve the phase prediction. A straightforward idea is to separately predict amplitude mask and phase with a two-stream network. However, Willianson et al. \cite{williamson2016complex} also pointed out that, in polar coordinates, structure does not exist in the phase spectrogram. This suggests that independent phase estimation is very difficult, if not completely impossible. In view of this, we add two-way information exchange for the two-stream architecture, so that the predicted amplitude can guide the prediction of phase. Results show that such information exchange is critical to the successful phase prediction of the target speech.

In the design of the amplitude stream, we find that conventional CNN kernels which are widely used in image processing do not capture the harmonics in T-F spectrogram. The reason is that correlations in natural images are mostly local while those in speech T-F spectrogram along the frequency axis are mostly non-local. In particular, at a given point of time, the value at a base frequency $f_0$ is strongly correlated with the values at its overtones. Unfortunately, previous DNN models cannot efficiently exploit harmonics although backbones like U-net \cite{jansson2017singing} and dilated 2D convolution \cite{ephrat2018looking} can increase the receptive field. In this paper, we propose to insert frequency transformation blocks (FTBs) to capture global correlations along the frequency axis. Visualization of FTB weights shows that FTBs spontaneously learn the correlations among harmonics.

In a nutshell, we design a \underline{p}hase-and-\underline{h}armonics-\underline{a}ware \underline{s}peech \underline{e}nhancement \underline{n}etwork, dubbed PHASEN, for monaural speech noise reduction. The contributions of this work are three-fold:
\begin{itemize}
    \item We propose a novel two-stream DNN architecture with two-way information exchange for efficient speech noise reduction in T-F domain. The proposed architecture is capable of recovering phase information of the target speech. 
    \item We design frequency transformation blocks in the amplitude stream to efficiently exploit global frequency correlations, especially the harmonic correlation in spectrogram. 
    \item We carry out comprehensive experiments to justify the design choices and to demonstrate the performance superiority of PHASEN over existing noise reduction methods.
\end{itemize}

The rest of this paper is organized as follows. Section \ref{sec:related} introduces related work. Section \ref{sec:arch} presents the proposed PHASEN architecture and its implementation details. Section \ref{sec:exp} shows the experimental results. Section \ref{sec:conclusion} concludes this paper with discussions on limitations and future work.

\begin{figure*}[t]
\centering
\includegraphics[width=\textwidth]{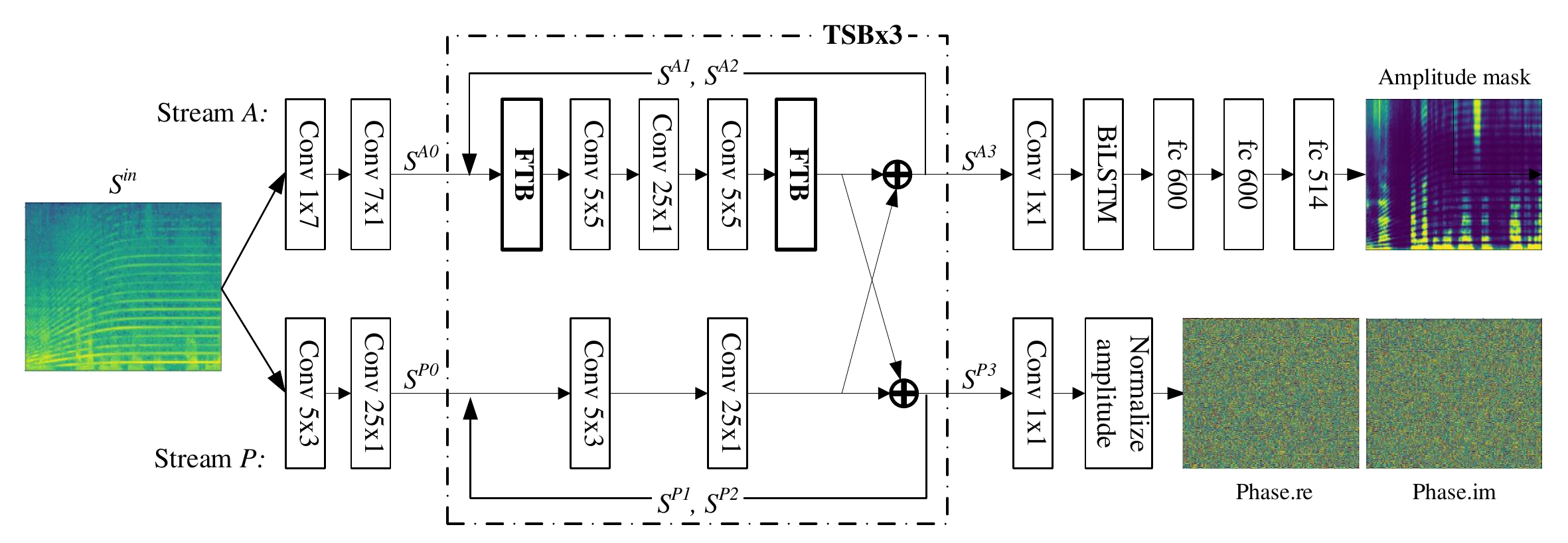} 
\caption{The proposed two-stream PHASEN architecture. The amplitude stream (Stream A) is in the upper portion and the phase stream (Stream P) is in the lower portion. The outputs of Stream A and Stream P are the amplitude mask and the estimated (complex) phase, respectively. Three two-stream blocks (TSBs) are stacked in the network. }
\label{fig:arch}
\end{figure*}

\section{Related Work}\label{sec:related}
This section reviews both time-frequency domain methods and time-domain methods for single-channel speech enhancement. Within T-F domain methods, we are only interested in T-F masking methods. Special emphases are put to phase estimation and the utilization of harmonics. 

\subsection{T-F Domain Masking Methods}
T-F domain masking methods for speech enhancement usually operate in three steps. First, the input time-domain waveform is transformed into T-F domain and represented by a T-F spectrogram. Second, a multiplicative mask is predicted based on the input spectrogram and applied to it. Last, an inverse transform is applied to the modified spectrogram to obtain the real-valued time-domain signal. The most widely used T-F spectrogram is computed by the short-time Fourier transform (STFT) and it can be convert back to time-domain signal by the inverse STFT (iSTFT). The key problems to be solved in T-F domain masking methods are what type of mask to be used and how to predict it. 

Early T-F masking methods only try to estimate the amplitudes of a spectrogram by using real-valued ideal binary mask (IBM) \cite{hu2001speech}, ideal ratio mask (IRM) \cite{srinivasan2006binary}, or spectral magnitude mask (SMM) \cite{wang2014training}. After the enhanced amplitudes are obtained, they are combined with the noisy phase to produce the enhanced speech. Later, research \cite{paliwal2011importance} reveals that phase plays an important role in speech quality and intelligibility. In order to recover phase, phase sensitive mask (PSM) \cite{erdogan2015phase} and cIRM \cite{williamson2016complex} are proposed. PSM is still a real-valued mask, extending SMM by simply adding a phase measure. In contrast, cIRM is a complex-valued mask which has the potential to faithfully recover both amplitude and phase of the clean speech. 


Williamson et al. \cite{williamson2016complex} propose a DNN-based approach to estimate the real and imaginary components of the cIRM, so that both amplitude
and phase spectra can be simultaneously enhanced. However, their experimental results show that using cIRM does not achieve significantly better results than using PSM. We believe that the potential of a complex mask is not fully exploited. In \cite{ephrat2018looking}, a much deeper neural network with dilated convolution and bi-LSTM is employed for speech separation with visual clues. It also achieves state-of-the-art speech enhancement performance when visual clues are absent. We carry out experiments on the network and surprisingly find that the imaginary components of the estimated cIRM is almost zero. This suggests that directly using cIRM to supervise a single-stream DNN cannot achieve satisfactory results.

There exist some other methods \cite{takahashi2018phasenet,takamichi2018phase,masuyama2019deep} which process phase reconstruction asynchronously with amplitude estimation. Their goal is to reconstruct phase based on a given amplitude spectrogram, which could be the amplitude spectrogram of a clean speech or the output from any speech denoising model. In particular, Takahashi et al. \cite{takahashi2018phasenet} observe the difficulty in phase regression, so they treat the phase estimation problem as a classification problem by discretizing phase values and assigning class indices to them. While all these methods demonstrate the benefits of phase reconstruction, their approach does not fully utilize the rich information in the input noisy phase spectrogram. 

\subsection{Time Domain Methods}
Time domain methods belong to the other camp for speech enhancement. We briefly mention several pieces of work here because they are proposed to avoid the phase prediction problem in T-F domain methods. SEGAN \cite{pascual2017segan} uses generative adversarial networks (GANs) to directly predict the 1D waveform of the clean speech. Rethage et al. \cite{rethage2018wavenet} modify Wavenet for the speech enhancement task. 
convolution-TasNet \cite{luo2019conv} uses a learnable encoder-decoder in time domain as an alternative to the hand-crafted STFT-iSTFT for a speech separation task. However, when it is applied to the speech enhancement task, the 2ms frame length appears to be too short. TCNN \cite{pandey2019tcnn} adopts a similar approach as TasNet, but it uses non-linear encoder-decoder and longer frame length than TasNet. Although these methods divert around the difficult phase estimation problem, they also give up the benefits of speech enhancement in T-F domain, as it is widely recognized that most speech and noise patterns are separately distributed or easily distinguishable on T-F domain features. As a result, the performance of time domain methods is not among the first tier in the speech enhancement task.

\subsection{Harmonics in Spectrogram}
Plapous et al. \cite{plapous2005speech} discover that common noise reduction algorithms suppress some harmonics existing in the original signal and then the enhanced signal sounds degraded. They propose to regenerate the distorted speech frequency bands by taking into account the harmonic characteristic of speech. Other research \cite{krawczyk2014stft,mowlaee2015harmonic} also show that phase correlation between harmonics can be used for speech phase reconstruction. A recent work \cite{wakabayashi2018single} further propose a phase reconstruction method based on harmonic enhancement using the fundamental frequency and phase distortion feature. All these work demonstrate the importance of harmonics in speech enhancement. In this paper, we also try to exploit harmonic correlation, but this is achieved by designing an integral block in the end-to-end learning DNN.


\section{PHASEN Architecture}\label{sec:arch}
\subsection{Overview}

The basic idea behind PHASEN is to separate the predictions of amplitude and phase, as the two prediction tasks may need different features. In our design, we use two parallel streams, denoted by stream $A$ for amplitude mask prediction and stream $P$ for phase prediction. The entire PHASEN architecture is shown in Fig. \ref{fig:arch}. 

The input to the network is the STFT spectrogram, denoted by $S^{in}$. Here, $S^{in} \in \mathbb{R}^{T\times F\times2}$ is a complex-valued spectrogram, where $T$ represents the number of time steps and $F$ represents the number of frequency bands. $S^{in}$ is fed into both streams and two different groups of 2D convolutional layers are used to produce feature $S^{A_0} \in \mathbb{R}^{T\times F \times C_{A}}$ for stream $A$ and feature $S^{P_0} \in \mathbb{R}^{T\times F \times C_{P}}$ for stream $P$. Here, $C_A$ and $C_P$ are the number of channels for stream $A$ and stream $P$, respectively.

The key component in PHASEN is the stacked two-stream blocks (TSBs), in which stream $A$ and stream $P$ features are computed separately. Note that at the end of each TSB, stream $A$ and stream $P$ exchange information. This design is critical to the phase estimation, as phase itself does not have structure and is hard to estimate \cite{williamson2016complex}. However, with the information from the amplitude stream, the features for phase estimation is significantly improved. In Section \ref{sec:exp}, we will visualize the difference between the estimated phase spectrograms when the information communication is present and absent. The output features of the three TSBs are denoted by $S^{A_i}$ and $S^{P_i}$, for $i \in \{1, 2, 3\}$. They have the same dimensions as $S^{A_0}$ and $S^{P_0}$. In stream $A$, frequency transformation blocks (FTBs) are used to capture non-local correlation along the frequency axis. 

After the three TSBs, $S^{A_{3}}$ and $S^{P_{3}}$ are used to predict amplitude mask and phase. For $S^{A_{3}}$, channel is reduced to $C_{r}=8$ by a $1\times 1$ convolution, then reshaped into a 1D feature map, whose dimension is $T\times (F \cdot C_{r})$, and finally fed into a Bi-LSTM and three fully connected (FC) layers to predict an amplitude mask $M \in \mathbb{R}^{T\times F \times 1}$. Sigmoid is used as activation function of the last FC layer. For the other FC layers, ReLU is used as activation function.

For $S^{P_{3}}$, a $1\times 1$ convolution is used to reduce channel number to 2 to form a complex-valued feature map $S^{P_{c}} \in \mathbb{R}^{T\times F \times 2}$, where the two channels correspond to the real and the imaginary parts. Then, amplitude of this complex feature map is normalized to 1 for each T-F bin. As such, the feature map only contains phase information. The phase prediction result is denoted by $\Psi$.

Finally, the predicted spectrogram can be computed by:
\begin{align}
S^{out} = abs(S^{in}) \circ M \circ \Psi,
\end{align}
where $\circ$ denotes element-wise multiplication.

\subsection{Two-Stream Blocks (TSBs)}

\subsubsection{Stream $A$}
In each TSB, three 2D convolutional layers are used for stream $A$ to handle local time-frequency correlation of the input feature. To capture global correlation on frequency axis such as harmonic correlation, we propose frequency transformation blocks (FTBs) to be used before and after the three convolutional layers. The FTB design will be detailed in the next subsection. The combination of 2D convolutions and FTBs efficiently captures both global and local correlations, allowing the following blocks to extract high-level features for amplitude prediction.
Stream $A$ of each TSB performs the following computation:
\begin{align}
S^{A_i}_{0} &= FTB_{in}^{i}(S^{A_i}), \\
S^{A_i}_{j+1} &= conv^{A_i}_{j}(S^{A_i}_{j}), \;\; j \in \{0, 1, 2\}, \\
S^{A_i}_{4} &= FTB_{out}^{i}(S^{A_i}_{3}). \label{equ:strmA}
\end{align}
Here, $conv^{A_i}_{j}$ represents the $j$-th convolutional layer in stream $A$ of the $i$-th TSB. $S^{A_i}_{j+1}$ and $S^{A_i}_{j}$ represent its output and input, respectively. $FTB_{in}^{i}$ and $FTB_{out}^{i}$ represent the FTB before and after the three 2D convolutional layers. Each 2D convolutional layer is followed by batch normalization (BN) and activation function ReLU. 

\subsubsection{Stream $P$}
Stream $P$ is designed to be light-weight. We only use two 2D convolutional layers in each TSB to process the input feature $S^{P_i} (i=1,2,3)$. Mathematically, 
\begin{align}  
S^{P_i}_{0} & = S^{P_i}, \\
S^{P_i}_{j+1} & = conv^{P_i}_{j}(S^{P_i}_{j}), \;\; for \;\; j \in \{0, 1\}. \label{equ:strmP} 
\end{align}  

Here, $conv^{P_i}_{j}$ represents the $j$-th convolutional layer in stream $P$ of the $i$-th TSB. $S^{P_i}_{j+1}$ and $S^{P_i}_{j}$ denote its output and input, respectively. The second convolutional layer uses a kernel size of 25$\times$1 to capture long-range time-domain correlation. Global Layer Normalization(gLN) is performed before each convolutional layer. In stream $P$, no activation function is used. We will later show in ablation studies that this choice increases performance. 


\subsubsection{Information Communication}
Information communication is critical to the success of the two-stream structure. Without the information from Stream $A$, Stream $P$ by itself cannot successfully make phase prediction. Conversely, successfully predicted phases can also help Stream $A$ to better predict amplitude. The communication takes place just before TSB generates output features. Let $S^{A_i}_{4}$ and $S^{P_i}_{2}$ be the amplitude features and phase features computed from eq. (\ref{equ:strmA}) and eq. (\ref{equ:strmP}), the output feature of TSB after information communication can be written as:
\begin{align}
S^{A_{i+1}} &= f_{P2A}(S^{A_i}_{4}, S^{P_i}_{2}), \\
S^{P_{i+1}} &= f_{A2P}(S^{P_i}_{2}, S^{A_i}_{4}),
\end{align}
where $f_{P2A}$ and $f_{A2P}$ are information communication functions of the two directions. In this work, we adopt the attention mechanism. For $i \in \{P2A, A2P\}$, we have:
\begin{align}
f_i(x_1, x_2) = x_1 \circ Tanh(conv(x_2)).
\end{align}

Here, $\circ$ denotes element-wise multiplication and $conv$ represents a $1\times 1$ convolution. The number of output channels is the same as the number of channels in $x_1$.


\begin{figure}[t]
\centering
\includegraphics[width=0.8\columnwidth]{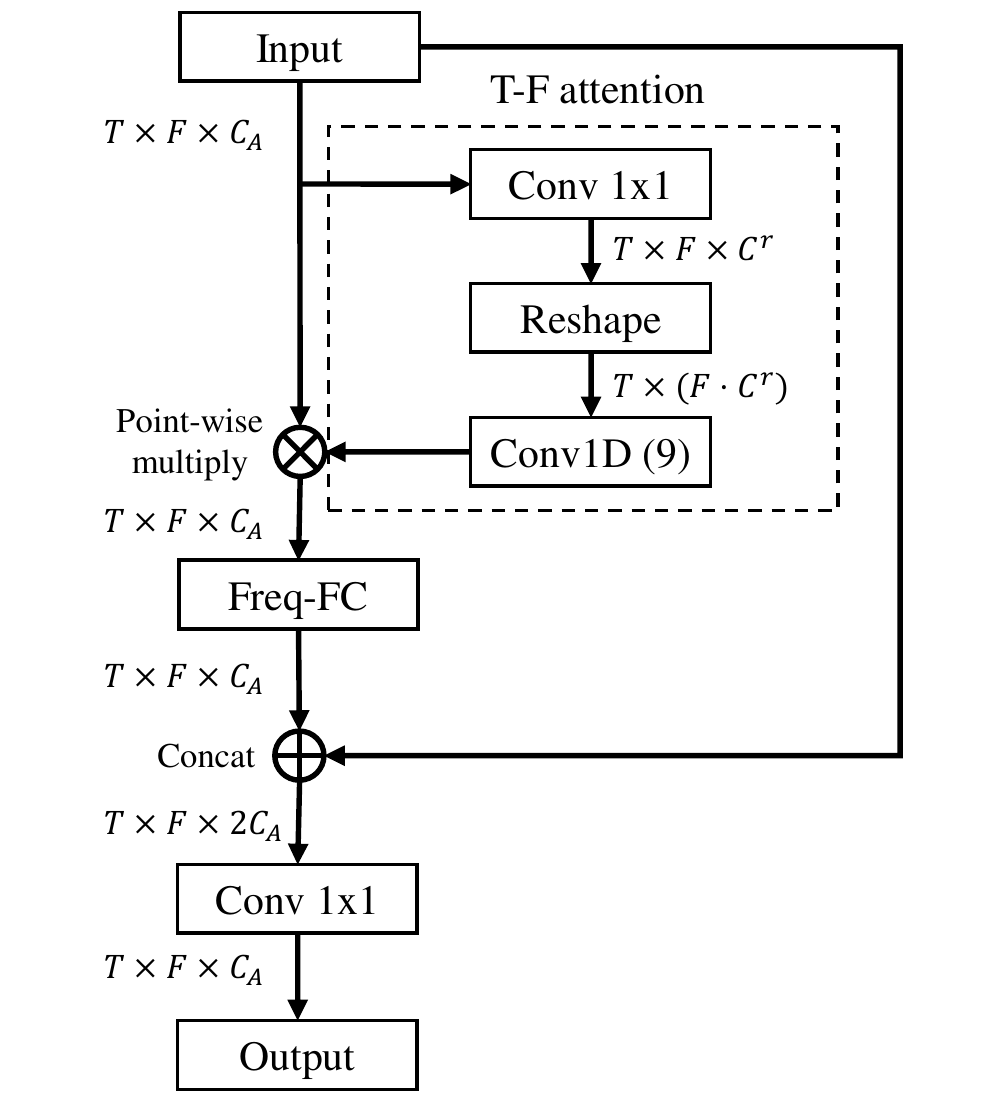} 
\caption{Flowchart of the proposed FTBs. Here, $C^r = 5$, and the kernel size of Conv 1D is 9.}
\label{fig6}
\end{figure}

\subsection{Frequency Transformation Blocks (FTBs)}
Non-local correlations exist in a T-F spectrogram along the frequency axis. A typical example is the correlations among harmonics, which has been shown to be helpful for the reconstruction of corrupted T-F spectrograms. However, simply stacking several 2D convolution layers with small kernels cannot capture such global correlation. Therefore, we design FTBs to be inserted at the beginning and the end of each TSB, so that the output features of TSB have full-frequency receptive field. At the kernel of an FTB is the learning of a transformation matrix, which is applied on the frequency axis. Fig. \ref{fig6} shows the flowchart of the proposed FTB. The three groups of operations in each FTB can be represented by:
\begin{align}
S^a &= f_{attn}(S^I), \\
S^{tr} &= \textrm{FreqFC}(S^a), \\
S^{O} &= conv(concat(S^{tr}, S^I)).
\end{align}

Eq. (10) describes the T-F attention module as highlighted in the dotted box in Fig. \ref{fig6}. With the input feature $S^I$, it uses 2D and 1D convolutional layers to predict an attention map, which is then point-wise multiplied to $S^I$ to obtain $S^a$. The 2D $1\times 1$ convolution reduces the channel number to $C^r=5$ and the kernel size of the 1D convolution is 9. 

Freq-FC is the key component in FTB. It contains a trainable frequency transformation matrix (FTM) which is applied to the feature map slice at each point in time. Let $X_{tr} \in \mathbb{R}^{F \times F}$ denote the trainable FTM and let $S^a(t_0) \in \mathbb{R}^{F \times C_A}$ $\left( t_0 \in \{0,1,...,T-1\} \right)$ denote the feature slice at time step $t_0$. The transformation can be simply represented by the following equation:
\begin{align}
S^{tr}(t_0) = X_{tr}\cdot S^a(t_0).
\end{align}
The transformed feature slice at time step $t_0$, denoted by $S^{tr}(t_0)$, has the same dimension as $S^a(t_0)$. Stacking them along the time axis and we can get the transformed feature map $S^{tr}$. After Freq-FC, each T-F bin in $S^{tr}$ will contain the information from all the frequency bands of $S^{a}$. This allows the following blocks to exploit global frequency correlations for amplitude and phase estimation.

The output of an FTB, denoted by $S^{O}$, is calculated by concatenating $S^{tr}$ with $S^I$ and fusing them with a $1\times 1$ convolution. In the proposed FTBs, batch normalization (BN) and ReLU are used after all convolutional layers as normalization method and activation function. 


\subsection{Implementation}
PHASEN is implemented in Pytorch. The dimension of feature maps and the kernel size of convolutional layers are shown in Fig. \ref{fig:arch} and Fig. \ref{fig6}. Both streams use convolution operation with zero padding, dilation=1 and stride=1, making sure the input and output feature map size are the same. All audios are resampled to 16kHz. STFT is calculated using Hann window, whose window length is 25ms. The hop length is 10ms and FFT size is 512. 

The network is trained using MSE loss on the power-law compressed STFT spectrogram. The loss consists of two parts: amplitude loss $L_a$ and phase-aware loss $L_p$.
\begin{align}
L &= 0.5\times L_a + 0.5\times L_p, \\
L_a &= MSE(abs(S^{out}_{cprs}), abs(S^{gt}_{cprs})), \\
L_p &= MSE(S^{out}_{cprs}, S^{gt}_{cprs}),
\end{align}
where $S^{out}_{cprs}$ and $S^{gt}_{cprs}$ are the power-law compressed spectrogram of output spectrogram $S^{out}$ and ground truth spectrogram $S^{gt}$. The compression is performed on amplitude with $p = 0.3$ ($A^{0.3}$, where $A$ is the amplitude of the spectrogram.) 

Note that instead of only using pure phase, whole spectrogram (phase and amplitude) is taken into consideration for $L_p$. In this way, phase of T-F bins with higher amplitude is emphasized, helping the network to focus on the high amplitude T-F bins where most speech signals are located.

\section{Experiments}\label{sec:exp}

\subsection{Datasets}\label{subsec:data}
Two datasets are used in our experiments.

\textbf{AVSpeech+AudioSet:} This is a large dataset proposed by \cite{ephrat2018looking}. Audios from AVSpeech dataset are used as clean speech. It is collected from YouTube, containing 4700 hours of video segments with approximately 150,000 distinct speakers, spanning a wide variety of people and languages. The noisy speech is a mixture of the above clean speech segments with AudioSet \cite{gemmeke2017audio}, which contains a total of more than 1.7 million 10-second segments of 526 kinds of noise. The noisy speech is synthesized by a weighted linear combination of speech segments and noise segments: $Mix_i = Speech_j + 0.3\times Noise_k$, where $Speech_j$ and $Noise_k$ are 3-second segments randomly sampled from speech and noise dataset. $Mix_i$ and $speech_j$ form a noisy-clean speech pair. In our experiments, 100k segments randomly sampled from AVSpeech dataset and the ``Balanced Train'' part of AudioSet are used to synthesize the training set, while the validation set is the same as the one used in \cite{ephrat2018looking}, synthesized by the test part of AVSpeech dataset and the evaluation part of AudioSet.

\textbf{Voice Bank+DEMAND:} This is an open dataset\footnote{https://datashare.is.ed.ac.uk/handle/10283/1942} proposed by \cite{valentini2016investigating}. Speech of 30 speakers from the Voice Bank corpus \cite{ephrat2018looking} are selected as clean speech: 28 are included in the training set and 2 are in the validation set. The noisy speech is synthesized using a mixture of clean speech with noise from Diverse Environments Multichannel Acoustic Noise Database (DEMAND) \cite{thiemann2013diverse}. A total of 40 different noise conditions are considered in training set and 20 different conditions are considered in test set. Finally, the training and test set contain 11572 and 824 noisy-clean speech pairs, respectively. Both speakers and noise conditions in the test set are totally unseen by the training set. Our system comparison is partly done on this dataset.

\subsection{Evaluation Metrics}
The following six metrics are used to evaluate PHASEN and state-of-the-art competitors. All these metrics are better if higher.
\begin{itemize}
  \item SDR \cite{vincent2006performance}: Signal-to-distortion ratio from the mir\_eval library;
  \item PESQ: Perceptual evaluation of speech quality (from -0.5 to 4.5).
  \item CSIG \cite{hu2007evaluation}: Mean opinion score (MOS) prediction of the signal distortion attending only to the speech signal (from 1 to 5).
  \item CBAK \cite{hu2007evaluation}: MOS prediction of the intrusiveness of background noise (from 1 to 5).
  \item COVL \cite{hu2007evaluation}: MOS prediction of the overall effect (from 1 to 5).
  \item SSNR: Segmental SNR .
\end{itemize}

\subsection{Ablation Study}
In the ablation study, networks of different settings are trained with the same random seed for 1 million steps. Adam optimizer with a fixed learning rate of 0.0002 is used and the batch size is set to 8. We use mean SDR and PESQ on test dataset as the evaluation metric. 

The ablation results are shown in Table \ref{tab:ablation}. Among these methods, PHASEN represents our full model. PHASEN-baseline represents a single-stream network which uses cIRM as training target. We use the network structure in stream $A$ for PHASEN-baseline and replace the FTBs with 5$\times$5 convolutions. The comparison between PHASEN and PHASEN-baseline shows that our two innovations, namely two-stream architecture and FTBs, provide a total of 1.76dB improvement on SDR and 0.53 improvement on PESQ.

\begin{table}[tb]
  \centering\small
  \caption{Ablation study on AVSpeech + AudioSet}
  \label{tab:ablation}
  \begin{tabular}{lcc}
    \hline
    Method & SDR(dB) & PESQ \\
    \hline
    PHASEN-baseline & 15.08 & 2.87 \\
    \hline
    PHASEN-1strm & 15.99 & 2.98 \\
    \hline
    PHASEN-w/o-FTBs & 16.10 & 3.31 \\
    \hline
    PHASEN-w/o-A2PP2A & 16.13 & 3.33 \\
    PHASEN-w/o-P2A & 16.62 & 3.38 \\
    \hline
    PHASEN & \textbf{16.84} & \textbf{3.40} \\
    \hline
  \end{tabular}
\end{table}

\subsubsection{Two-Stream Architecture}
PHASEN-1strm shows the performance of single-stream architecture with cIRM as training target. In this experiment, stream $P$ and information communication are removed from PHASEN architecture, while FTBs are preserved. The output of stream $A$ is the predicted cRM. Comparison between PHASEN-1strm and PHASEN shows that the two-stream architecture provides 0.85dB gain on SDR and 0.42 gain on PESQ. The large gain on PESQ indicates the proposed two-stream architecture can largely improve the perceptual quality of the denoised speech.

\subsubsection{FTBs}
The proposed method uses FTBs at both the beginning and the end of each TSB. In ablation study, PHASEN-w/o-FTBs try to replace all the FTBs in PHASEN architecture with 5$\times$5 convolutions. By comparing PHASEN to PHASEN-w/o-FTBs we find that FTBs can provide 0.74 dB and 0.09 gain on SDR and PESQ, respectively. We have also tried to replace the FTBs on either location of each TSB with 5$\times$5 convolutions. Both attempts result into 0.31dB-0.39dB drop on SDR and 0.03-0.05 drop on PESQ, showing that FTBs on both locations are equally important and the gain is accumulative. 

\begin{figure}[t]
\centering
\includegraphics[width=\columnwidth]{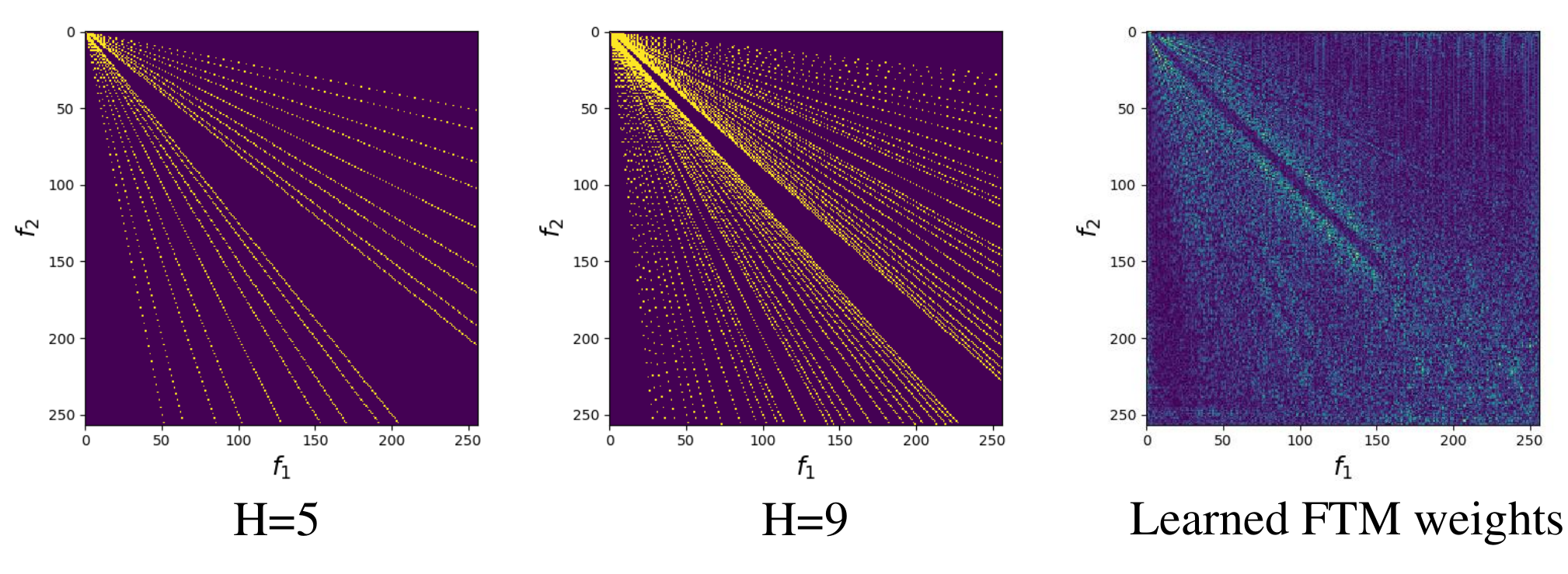} 
\caption{Comparison of different level of harmonic correlation: $f_2 = \frac{m}{n}f_1, m\neq n, m,n \in \{0,1,...,H\}$ and learned FTM weights. $f_1=f_2=0$ is on the upper-left corner of each sub-figure.}
\label{fig7}
\end{figure}

In order for a better understanding of FTBs, we visualize the weights of $X_{tr}$, the matrix that reflects the learned global frequency correlation. From Fig. \ref{fig7}, we show that the energy map of $X_{tr}$ resembles the harmonic correlation, especially when higher harmonics (larger H) are taken into consideration. This phenomenon confirms that FTBs really capture the harmonic correlation, and that harmonic correlation is really useful to a speech enhancement network, because the network can learn this correlation spontaneously.

\subsubsection{Information communication mechanism}
PHASEN-w/o-P2A, and PHASEN-w/o-A2PP2A are two settings that remove the information communication mechanism partly and fully. The former one removes the communication from stream $P$ to stream $A$, and the latter one removes communication of both directions. In SDR and PESQ result, significant gain of 0.49dB and 0.05 is observed when comparing PHASEN-w/o-P2A to PHASEN-w/o-A2PP2A. This indicates that the information in the intermediate steps of amplitude prediction is very helpful to phase prediction. In comparison between our full model PHASEN and PHASEN-w/o-P2A, we also see that when integrating stream $P$ information into stream $A$, the model gets 0.22dB gain on SDR and 0.02 gain on PESQ. This proves that phase feature can also help amplitude prediction.

\begin{figure}[t]
\centering
\includegraphics[width=\columnwidth]{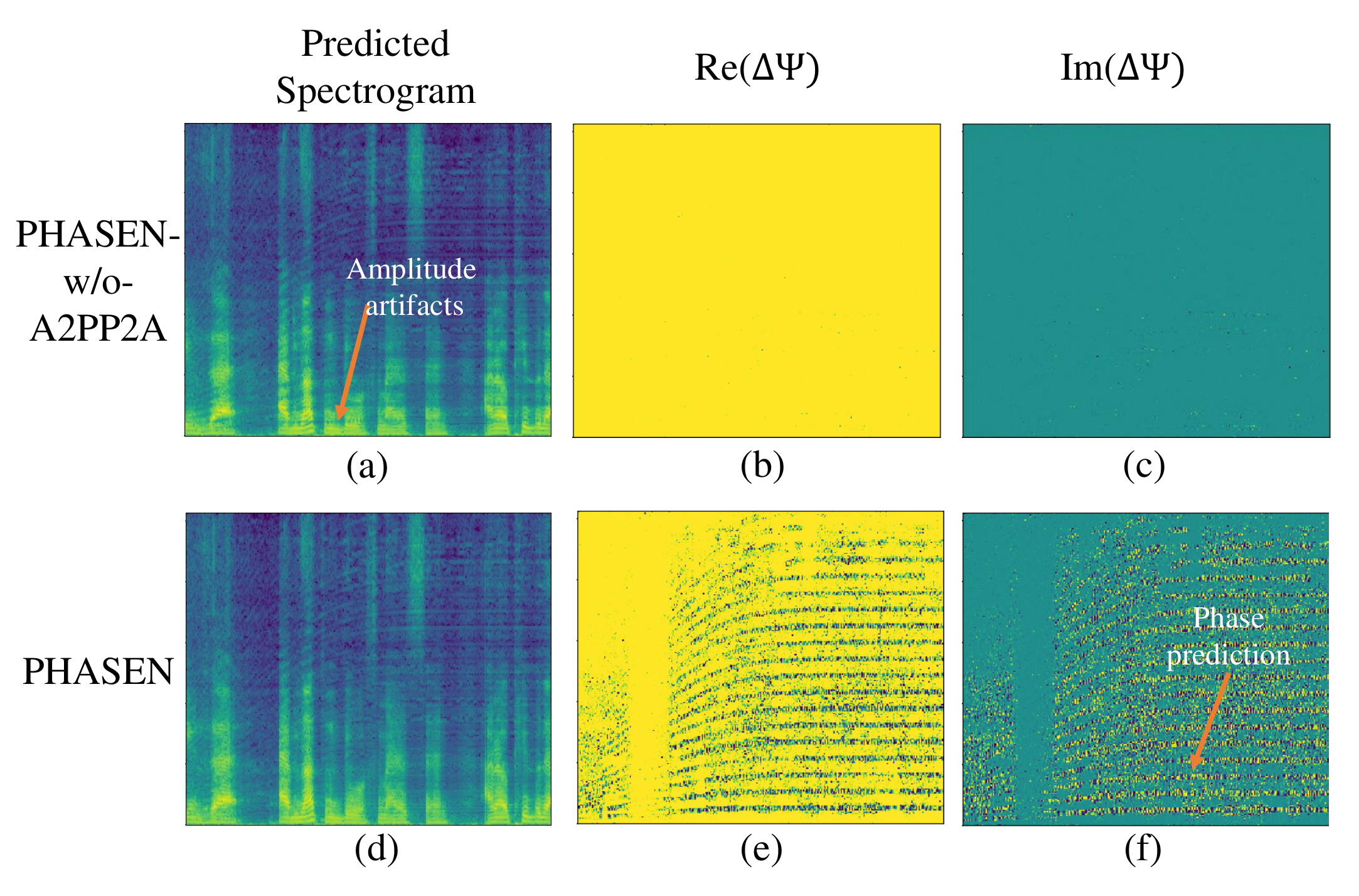} 
\caption{The effect of information communication mechanism. Best viewed in color. We use the same input noisy speech as the one in Fig.\ref{fig:arch} to produce the results. (a),(b),(c): Amplitude of predicted spectrogram, real part, and imaginary  part of $\Delta\Psi$ in setting PHASEN-w/o-A2PP2A. Significant amplitude artifacts are observed in (a) on frequency bands where speech is overwhelmed by noise. In every T-F bins, (c) is almost zero, and (b) is almost one, indicating failure on phase prediction. (d),(e),(f): Amplitude of predicted spectrogram, real part, and imaginary part of $\Delta \Psi$ in setting PHASEN. In (e) and (f), phase prediction is obviously visible in T-F bins where noise overwhelms speech. (d) also shows fewer artifacts on amplitude spectrogram.}
\label{fig8}
\end{figure}

Fig. \ref{fig8} also confirms the above improvements through visualization. Here, because the predicted phase spectrogram has few visible patterns, we visualize $\Delta\Psi = \Psi / \Psi_{in}$, which represents the phase difference between predicted phase spectrogram and input noisy spectrogram. The division operation in this formula is on complex domain, and $\Psi_{in}$ represents the phase spectrogram of input noisy speech. From the visualization, we can conclude that information communication mechanism not only significantly improves the phase prediction, but helps remove amplitude artifacts. 
To summarize, information communication of both directions are useful in PHASEN, while direction ``A2P'' plays a key role.


\subsubsection{Other ablations}
Apart from the results shown in Table \ref{tab:ablation}, we also perform ablations on activation function and normalization functions for stream $P$.

The proposed method uses no activation function on stream $P$. Though this design is counter-intuitive, it is actually inspired by previous work \cite{luo2019conv} and also supported by the ablation study. In fact, we try to add ReLU or Tanh as activation function after each, except the last, convolutional layer in stream $P$. However, this causes 0.02dB-0.16dB drop on SDR. Moreover, if ReLU is added after the last convolutional layer in stream $P$, a huge drop of 5.52dB and 0.2 is observed on SDR and PESQ. 

The proposed method uses gLN in stream $P$ and BN in stream $A$. We test other normalization method for each stream. A performance drop of 0.97dB and 0.12 on SDR and PESQ is observed if gLN is used in stream $A$, while a drop of 0.09dB and 0.02 on SDR and PESQ is observed if BN is used in stream $P$.

From these two experiments, we can observe significant difference between phase prediction and amplitude mask prediction. This supports our design of using two streams to accomplish the two prediction tasks.

\subsection{System Comparison}
We carry out system comparison on both datasets mentioned in section \ref{subsec:data}.

\subsubsection{AVSpeech + AudioSet} 
On this large dataset we compare our method with two other recent methods, Conv-TasNet \cite{luo2019conv} and ``Google'' \cite{ephrat2018looking}. Conv-TasNet is a time domain method. The result of Conv-TasNet is produced using the released code\footnote{https://github.com/funcwj/conv-tasnet}, trained for the same epochs and on the same data as our PHASEN. ``Google'' is a T-F domain masking method which uses cIRM as supervision. The method is intended for both speech noise reduction and speech separation. We compare PHASEN with their audio-only, 1S+noise setting. The result in Table \ref{table4} shows that our method outperforms both Conv-TasNet and ``Google''. Note that this is achieved under the condition that we only use a small fraction of training step (1M/5M) and data (100k/2.4M) used by ``Google''. Such superior performance on large dataset demonstrates that our method can be generalized to various speakers and various kinds of noisy environments. It suggests that PHASEN is readily applicable to complicated real-world environment.

\begin{table}[tb]
  \centering\small
  \caption{System comparison on AVSpeech + AudioSet}
  \label{table4}
  \begin{tabular}{lcc}
    \hline
    Method & SDR(dB) & PESQ \\
    \hline
    Conv-TasNet & 14.19 & 2.93 \\
    Google(5M step, 2.4M speech) & 16.00 & -- \\
    PHASEN(1M step, 100k speech) & \textbf{16.84} & \textbf{3.40} \\
    \hline
  \end{tabular}
\end{table}

\subsubsection{Voice Bank + DEMAND}
Apart from using large dataset, we also train our model on small but commonly-used dataset Voice Bank + DEMAND, so that we can fairly compare our PHASEN with many other methods. In this experiment, our network is trained on training set for 40 epochs, with Adam optimizer using warm-up step number of 6000, learning rate of 0.0005, and batch size of 12. 

Table \ref{table5} shows the comparison result. Firstly, our method has very large gain over time-domain methods like SEGAN \cite{pascual2017segan}, Wavenet \cite{rethage2018wavenet}, and DFL \cite{germain2018speech} on all the five metrics, even though these time-domain methods are free of phase-prediction problem. This proves the advantage of our method over the time-domain methods on capturing phase-related information. Also, our method shows great improvement over time-frequency domain method like MMSE-GAN \cite{soni2018time} on all metrics, indicating the superiority of our network design. Finally, we also compare our method with a recent hybrid model of time-domain and time-frequency domain called MDPhD \cite{kim2018multi}. Our method significantly outperforms it on four metrics, and there is only a small difference of about 0.04dB on SSNR metric.

\begin{table}[tb]
  \centering\small
  \caption{System comparison on Voice Bank + DEMAND}
  \label{table5}
  \begin{tabular}{lccccc}
    \hline
    Method & SSNR & PESQ & CSIG & CBAK & COVL \\
    \hline
    Noisy & 1.68 & 1.97 & 3.35 & 2.44 & 2.63\\
    \hline
    SEGAN & 7.73 & 2.16 & 3.48 & 2.94 & 2.80\\
    Wavenet & -- & -- & 3.62 & 3.23 & 2.98\\
    DFL & -- & -- & 3.86 & 3.33 & 3.22\\
    \hline
    MMSE-GAN & -- & 2.53 & 3.80 & 3.12 & 3.14\\
    \hline
    MDPhD & \textbf{10.22} & 2.70 & 3.85 & 3.39 & 3.27\\
    \hline
    PHASEN & 10.18 & \textbf{2.99} & \textbf{4.21} & \textbf{3.55} & \textbf{3.62}\\
    \hline
  \end{tabular}
\end{table}

\section{Conclusion}\label{sec:conclusion}
We have proposed a two-stream architecture with two-way information communication for efficient phase prediction in monaural speech enhancement. We have also designed a learnable frequency transformation matrix in the network. It spontaneously learns a pattern that is consistent with harmonic correlation. Comprehensive ablation studies have been carried out, justifying almost every design choices we have made in PHASEN. Comparison with state-of-the-art systems on both AVSpeech+AudioSet and Voice Bank+DEMAND datasets demonstrates the superior performance of PHASEN. Note that the current design of PHASEN does not allow it to be used for low-latency applications, such as voice over IP. In the future, we plan to explore the potential of PHASEN in low-latency settings and mobile settings which require a smaller model size and shorter inference time. We also plan to expand this architecture to other related tasks such as speech separation.

\bibliographystyle{aaai}
\bibliography{reference}

\end{document}